\documentclass[superscriptaddress,twocolumn,showpacs,aps,prl]{revtex4-1}

\usepackage{amssymb,xcolor}
\usepackage{amsmath}
\usepackage{amsbsy}
\usepackage{amsthm}
\usepackage{graphicx}
\usepackage{amsfonts}
\usepackage{times}
\usepackage{txfonts}

\newcommand{\beq}{\begin{equation}}
\newcommand{\eeq}{\end{equation}}

\newcommand{\bra}[1]{\langle#1|}
\newcommand{\proj}[1]{|#1\rangle\langle#1|}
\newcommand{\ket}[1]{|#1\rangle}

\newcommand{\id}{\leavevmode\hbox{\small1\normalsize\kern-.33em1}}

\newcommand{\eq}[1]{Eq. (\ref{#1})}

\begin{document}

\title{Experimental Realization of Optimal Noise Estimation for a General
Pauli Channel}

\author{A. Chiuri}
\affiliation{Dipartimento di Fisica, Sapienza {Universit\`a} di Roma, Piazzale Aldo Moro 5, I-00185 Roma, Italy}
\affiliation{Istituto Nazionale di Ottica (INO-CNR), Largo Enrico Fermi 6, I-50125 Firenze, Italy}
\author{V. Rosati}
\affiliation{Dipartimento di Fisica, Sapienza {Universit\`a} di Roma, Piazzale Aldo Moro 5, I-00185 Roma, Italy}
\author{G. Vallone}
\affiliation{Dipartimento di Fisica, Sapienza {Universit\`a} di Roma, Piazzale Aldo Moro 5, I-00185 Roma, Italy}
\affiliation{Department of Information Engineering, University of Padova, I-35131 Padova, Italy}
\author{S. P\'adua}
\affiliation{Departamento de F\'{\i}sica, Universidade Federal de
Minas Gerais. Caixa Postal 702, Belo~Horizonte,~MG 30123-970,
Brazil}
\author{H. Imai}
\affiliation{Dipartimento di Fisica ``A. Volta'' and INFN-Sezione di Pavia, via Bassi 6, 27100 Pavia, Italy}
\author{S. Giacomini}
\affiliation{Dipartimento di Fisica, Sapienza {Universit\`a} di Roma, Piazzale Aldo Moro 5, I-00185 Roma, Italy}
\author{C. Macchiavello}
\affiliation{Dipartimento di Fisica ``A. Volta'' and INFN-Sezione di Pavia, via Bassi 6, 27100 Pavia, Italy}
\author{P. Mataloni}
\affiliation{Dipartimento di Fisica, Sapienza {Universit\`a} di Roma, Piazzale Aldo Moro 5, I-00185 Roma, Italy}
\affiliation{Istituto Nazionale di Ottica (INO-CNR), Largo Enrico Fermi 6, I-50125 Firenze, Italy}

\date{\today}

\begin{abstract}
We present the experimental realization of the optimal estimation protocol
for a Pauli noisy channel. The method is based on the generation of 2-qubit
Bell
states and the introduction of quantum noise in a controlled way on
one of the state subsystems.
The efficiency of the optimal estimation, achieved by a Bell
measurement, is shown to outperform quantum process tomography.
\end{abstract}


\maketitle

\begin{figure}[t]
\centering
\includegraphics[width=8cm]{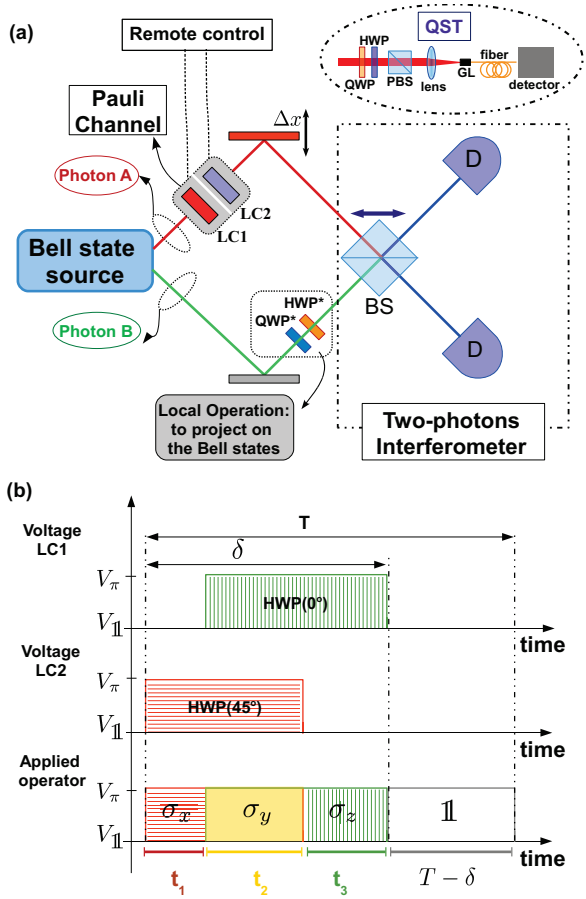}
\caption{ (a) Experimental setup. Photons A and B are spatially and temporally superimposed on a symmetric beam splitter (BS).
The optical path delay $\Delta x$ allows us to vary the arrival time of the photons on the BS.
Photons are collected by using an integrated system, composed by a GRIN lens (GL)
and a single mode fiber, and then detected by single photon counters.
The same setup allows us to perform the ancillary assisted quantum process tomography (AAQPT) after removing the BS.
Quantum state tomography (QST) \cite{jam01pra} on the output state is performed
by using quarter-wave plates (QWPs), half-wave plates (HWPs), and polarizing beam splitters (PBSs).
(b) Scheme of the
implemented Pauli channel.
$t_1$, $t_2$,$t_3$ represent the time intervals of $\sigma_x$, $\sigma_y $ or $\sigma_z$ activation. Both $t_1$, $t_2$, $t_3$ and the repetition time $T$
can be varied by a remote control.
}
\label{setup}
\end{figure}

{\it Introduction.- }
Quantum noise is unavoidably present in any realistic
implementation of quantum tasks, ranging from quantum communication
protocols \cite{cira99pra} to
quantum information processing devices and quantum metrology
\cite{giov04science,giov06prl}. The performance and the
optimization of quantum tasks quite often depend on the level of noise which
is present in the physical realization considered.
It is therefore
of great interest to develop experimental methods to estimate the level of
noise in the system under examination as precisely as possible.
Quantum process tomography (QPT) \cite{nielsen},
which has already been implemented in various experimental realizations
\cite{qpt-exp, obri04prl},
represents a well-known method to identify an unknown noise, but it
lacks the notion of efficiency. In many realistic scenarios, however, some
\textit{a priori} information on the kind
of noise is available and therefore the problem of measuring it is equivalent
to estimate few noise parameters in the most efficient way.
This is the context of quantum channel estimation \cite{qc-est},
which is based on quantum state estimation theory,
a merge of statistical estimation theory and quantum physics pioneered by
Helstrom and Holevo \cite{Helstrom,Holevo}.
The aim of this Letter is to provide
the first genuine experimental application of quantum channel estimation
theory. The experimental realization presented here is based on a quantum
optical setup, but it opens new perspectives of applications to a great
variety of physical scenarios and quantum technologies,
from atomic to solid state systems.

A quantum channel estimation problem is generally formulated as follows.
We need to estimate the true value of the $d$-dimensional parameter $\theta$
for a given smooth parametric family $\Gamma_\theta$ of noisy quantum
channels.
The estimation scheme is twofold: first we prepare a quantum
system described by the density operator $\rho$ as input to the channel
$\Gamma_\theta$,
then we perform some quantum state measurement on the output state
$\Gamma_\theta\left(\rho\right)$ in order to
estimate the channel parameters in the most efficient way.
Thus, the problem is to seek an optimal input $\rho$ for the channel
and an optimal measurement on the output state
$\Gamma_\theta\left(\rho\right)$.
The notion of optimality is here based on the
minimization of the covariance matrix $V_\theta[\rho,M]$
\begin{equation}
V_{\theta}[\rho,M]^{ij}={E}_{\theta}\left[(\check{\theta}^{i}-
\theta^{i})(\check{\theta}^{j}-\theta^{j})\right]\;,
\label{covmat}
\end{equation}
where $i,j=1,...,d$, $\check{\theta}^{j}$ and $\theta^{j}$
denote the estimated and the true values for the $j$-th component of the noise
parameter respectively, ${E}_{\theta}$ denotes the expectation
with respect to the measurement procedure employed, and $M$ is a projective measurement operator.

In this work we present an experimental implementation of an optimal
quantum channel estimation scheme for a qubit Pauli channel (PC).
The action of such a family of channels on the density operator $\rho$
of a qubit can be described as \cite{nielsen}
\begin{equation}\label{channel}
\Gamma_{\{p\}}[\rho]= {\sum^3}_{i=0}{p_i \sigma_i \rho \sigma_i}
\end{equation}
where $\sigma_0$ is the identity operator, \{$\sigma_i$\} ($i=1,2,3$)
are the three Pauli operators $\sigma_x ,\sigma_y, \sigma_z$ respectively, and
\{$p_i$\} represent the corresponding probabilities ($\sum^3_{i=0}p_i=1$).
The family of Pauli channels represents a wide class of noise
processes, that includes several physically relevant cases such as the
depolarising channel, which will be considered in the following, the
dephasing and the bit-flip channels.

The optimal channel estimation scheme is achieved as follows
\cite{opt-qce}. The optimal
input state is represented by a Bell state for two qubits,
for example the singlet state ${\ket{\psi^-}}= \frac{1}{\sqrt{2}}({\ket {01}}-{\ket {10}})$,
where only
one of the qubits is affected by the noisy channel while the other one is
left untouched. The optimal measurement consists of a Bell measurement
on the two qubits at the channel output, namely the projective
measurement
$M=\left\{\proj{\psi^-}, \proj{\psi^+},\proj{\phi^-},\proj{\phi^+}\right\}$.
The outcome probabilities then provide an optimal estimation of the channel
parameters $p_i$.

As mentioned above, this scheme is optimized by
minimizing the covariance matrix of the estimation error (\ref{covmat}).
According to the quantum Cram\'er-Rao theorem \cite{Holevo}, the
minimum covariance matrix in this case is given by \cite{opt-qce}
\begin{equation}
{V}_{p,min}=J_{p}[\rho_{ME}]^{-1}=\left[\begin{array}{ccc}
p_{1}(1-p_{1}) & -p_{2}p_{1} & -p_{3}p_{1}\\
-p_{1}p_{2} & p_{2}(1-p_{2}) & -p_{3}p_{2}\\
-p_{1}p_{3} & -p_{2}p_{3} & p_{3}(1-p_{3})\end{array}\right]
\end{equation}
where $J_{p}[\rho_{ME}]$ is the quantum Fisher information matrix
\cite{Holevo} of a maximally entangled input state $\rho_{ME}$.
We want to point out that this
scheme is optimal for any number of input qubits. Actually, no additional
entanglement among the input qubits and no collective measurements at the
output can increase the efficiency of the present scheme \cite{opt-qce}.
Moreover, it can be also straightforwardly generalized to estimate any
general noise process of the form (\ref{channel}), where the $\sigma$
operators are replaced by any set of unitary operators $V_i$ such that
Tr$[V_iV_j^\dagger]=2\delta_{ij}$.
The same scheme can also straightforwardly extended to estimate
any generalized Pauli channel for quantum systems in arbitrary finite
dimension \cite{opt-qce}.

We will now present the experimental implementation of this optimal
estimation scheme for
a quantum optical setup, where the state of the two qubits is represented
by polarization states of two photons and the action of the Pauli channel
is introduced in a controlled way by employing liquid crystal retarders,
as explained in the following.
The method has been first applied to estimate a general Pauli channel, with
independent values of the probabilities $p_i$.
Then it has been applied to a depolarizing channel (DC), namely the
case of isotropic noise, with $p_1=p_2=p_3=\frac{p}{3}$, where the parameter $p$ completely specifies
the channel itself,  and the minimum variance (Eq. (\ref{covmat}) for the
one-dimensional case) is given by $p(1-p)$. In this case the procedure simplifies
and, in the following, we show that only two projective measurements, $M'=\left\{\proj{\psi^-}, 1-\proj{\psi^-}\right\}$, are needed.

\begin{figure}[h!]
\centering
\includegraphics[angle=270,width=8cm]{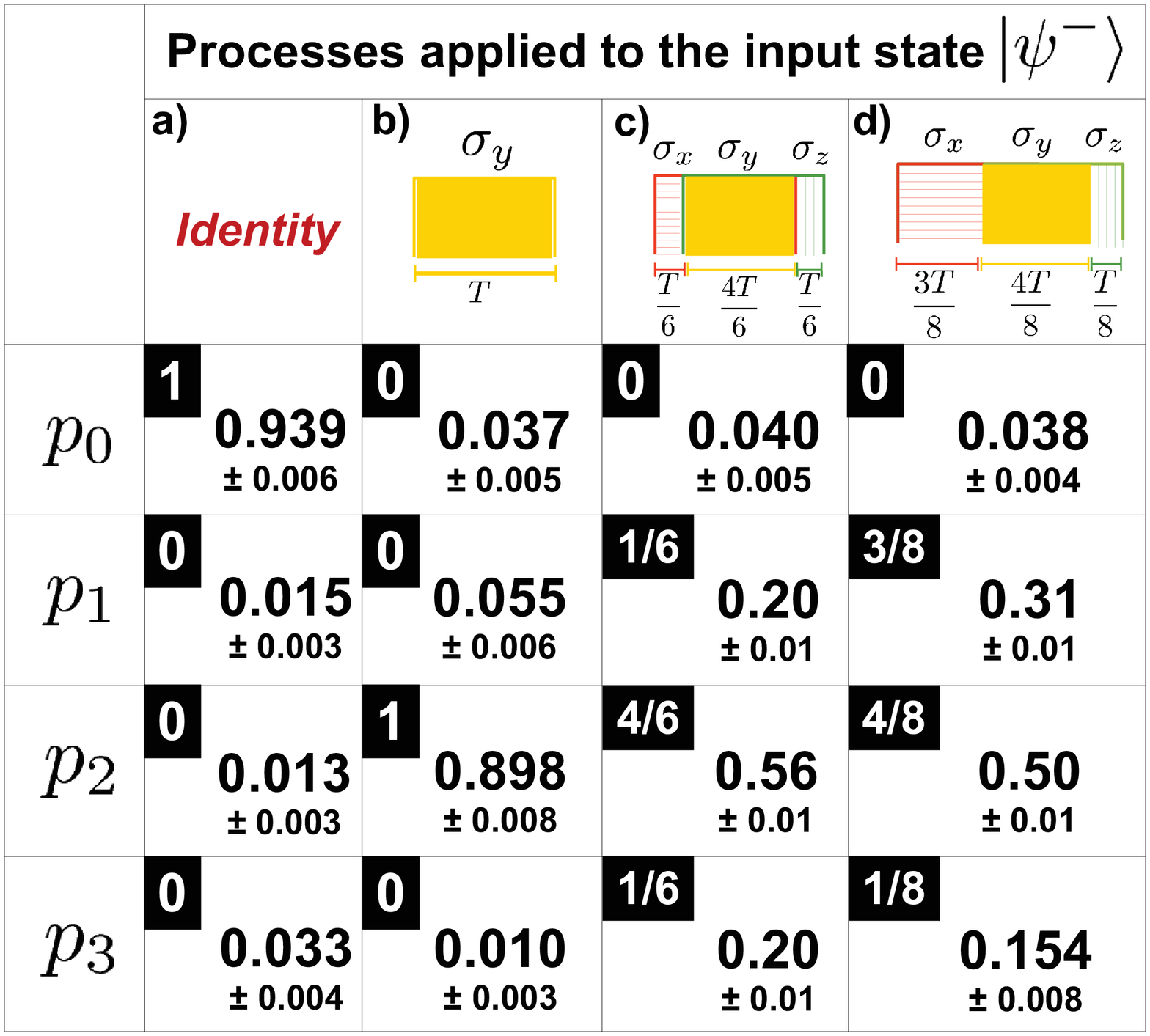}
\caption{Experimental probabilities of measuring the four Bell states obtained for four different cases of anisotropic noise.
The black boxes report the corresponding theoretical values.
a) \textit{Identity}: noiseless channel.
b) $\sigma_y $: only one Pauli matrix, $\sigma_y$ is acting on the state $\ket{\psi^-}$.
c) \textit{Partially anisotropic DC}: $\sigma_x$ and $\sigma_z$
operate for the same time interval, in fact the probabilities of measuring the states $\ket{\psi^+}$ and $\ket{\phi^+}$ are equal.
d) \textit{Totally anisotropic DC}: each Pauli operator operates for a different time interval.
}
\label{anisotropo}
\end{figure}

\begin{figure*}[t]
\includegraphics[angle=270,width=18cm]{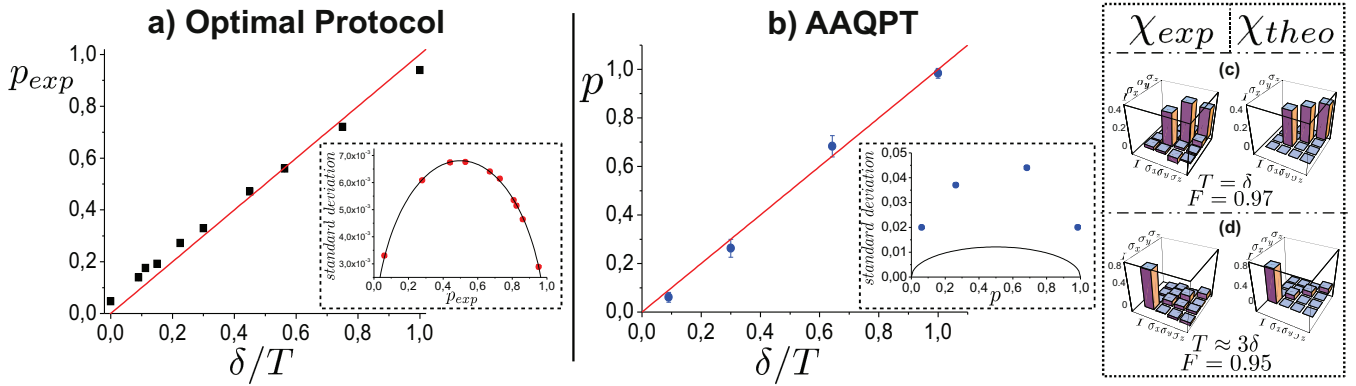}
\caption{Noise parameter estimation for the DC case.  
(a) Measured values of $p_{exp}$ vs $\frac{\delta}{T}$ by implementing the $M'$
projective measurements. Continuous red line corresponds to the theoretical behavior. 
Inset: experimental values of the standard deviations
for the optimal protocol implemented by the $M'$ projective measurements.
They are obtained by propagating the 
Poissonian uncertainties.  
The solid line
represents the expected theoretical behavior.
b): Experimental probabilities associated to the experimental matrix $\chi$ vs $\frac{\delta}{T}$.
Values of $p$ are obtained by maximizing the fidelity $F$ between theoretical and experimental matrix $\chi$.
Error bars are calculated by considering the Poissonian uncertainty
associated to the coincidence counts, and simulating different matrices of the process,
obtaining, in this way, different values of $p$. Inset: experimental values of the standard deviations for the AAQPT. 
These have been calculated by a simulation based on the poissonian uncertainty associated to the 
coincidence counts. The solid line represents the optimal bound.
(c),(d) Experimental (left side) and theoretical (right side) matrices $\chi$ for $T=\delta$ and $T\approx 3 \delta$.}
\label{probability}
\end{figure*}




{\it Experimental scheme.- }
Different techniques have been exploited
to experimentally implement a PC acting on a single qubit state \cite{jinprl09,ricprl04, karjosab04,eise11pra}.
The optimal noise estimation protocol, proposed in this work,
was implemented by the interferometric scheme shown in Fig.\ref{setup}a).
Precisely, a two-photon entangled source \cite{cinpra04} generates the two-qubit singlet state
${\ket{\psi^-}}= \frac{1}{\sqrt{2}}({\ket {HV}}_{AB}-{\ket {VH}}_{AB})$,
where two qubits are encoded in the polarization degree of freedom,
with $H$ ($V$) referring to the horizontal (vertical) polarization of photons
$A$ and $B$.
In our setup, the single qubit noisy channel is operating only on one of the two entangled particles (i.e. photon A).
The general Pauli channel (PC) consists of a sequence of liquid crystal retarders (LC1 and LC2) in the path of photon $A$.
The LCs act as phase retarders, with the relative phase between the ordinary and extraordinary
radiation components depending on the applied voltage $V$.
Precisely, $V_{\pi}$ and $V_{\id}$ (Fig.\ref{setup}b) correspond to the
case of LCs operating as half-wave plate (HWP) and as the
identity operator, respectively.
The LC1 and LC2 optical axes are set at $0^\circ$ and $45^\circ$ with respect to the V-polarization.
Then, when the voltage $V_{\pi}$ is applied, the LC1 (LC2) acts as a $\sigma_z$ ($\sigma_x$) on the single qubit.
We were able to switch between $V_{\id}$ and $V_\pi$
in a controlled way and independently for both LC1 and LC2.
The simultaneous application of $V_{\pi}$ on both LC1 and LC2 corresponds to the $\sigma_y $ operation.
We could also adjust the temporal delay
between the intervals in which the $V_\pi$ voltage is applied to the two retarders.
We define $t_1$, $t_2$, $t_3$ respectively as the activation time of the operators
$\sigma_x$, $\sigma_y $ or $\sigma_z$ and $T$  is the period of the LCs activation cycle, as shown in Fig.\ref{setup}b).

{\it Experimental implementation of the Pauli channel
(anisotropic noise).- }
A general PC was generated by varying the four time intervals $t_1$, $t_2$, $t_3$ and $T$.
The intervals $t_i$ are related to the probabilities $p_i$ ($i=1,2,3$), introduced in \eq{channel},
by the following expression: $p_i=\frac{t_i}{T}$.
The probability $p_0$ of the identity operator is given by $p_0=1-\frac\delta T$
(with $\delta=t_1+t_2+t_3$). 

To obtain the probabilities associated to the four projection operators $M$,
we measured the coincidence counts between
the two outputs of the BS. In fact, these probabilities
are related to the interference visibility measured by the interferometer in Fig.\ref{setup}a).
The half-wave plate (HWP$^*$) and quarter-waveplate (QWP$^*$) of Fig.\ref{setup}a) were used to project the noisy state
onto the four different Bell states.

Different configurations of the noisy channel
were investigated by implementing the optimal noise
protocol estimation for each configuration.
A summary of four relevant experimental results, corresponding to different probabilities
associated to the Bell states, are given in Fig.\ref{anisotropo}.

In the measurements shown in Fig.\ref{anisotropo}, case (a)
correspond to a noiseless channel (identity transformation) while the cases b), c) and d), correspond
to different complete noisy channels with $p_{0}=0$ (i.e. we set $T=\delta$).
For each process, the first column shows the relative weights between the Pauli operators acting in the channel.
From these values it is possible to calculate the theoretical ones. For instance,  let us consider the process d)
where the $\sigma_z$, $\sigma_y$ and $\sigma_x$ act respectively for $\frac{T}{8}$, $\frac{4T}{8}$ and $\frac{3T}{8}$.
The expected values of $p_i$ are, for this process, $p_0=0$, $p_1=\frac{3}{8}$,  $p_2=\frac{4}{8}$ and $p_3=\frac{1}{8}$.
The slight disagreement between
the expected theoretical values and the experimental measured ones are mainly due to the finite
rise and decay times of the electrical signal driving the LC devices.

We have implemented the protocol by using
always the same input state and projecting it on the Bell basis.
It is worth noting that this is totally equivalent to entering the
PC with the four Bell states and to projecting them
into the $\ket{\psi^-}$ state.

{\it Experimental implementation of the depolarizing channel (isotropic noise).- }
The condition $t_1=t_2=t_3$ corresponds to the depolarizing channel,
with the three Pauli operators acting on the single qubit with the same probability $p=\frac{\delta}{T}=\frac{t_1+t_2+t_3}{T}$.
This parameter was changed by fixing the times $t_i$ and varying the period $T$.
The optimal protocol to estimate the value of $p$ was realized by using the Bell state $\ket{\psi^-}$, as
mentioned above.

The DC was activated on photon $A$. In this case the projective measurement $M'=\{ \ket{\psi^-}\bra{\psi^-}, \id -\ket{\psi^-}\bra{\psi^-}\}$, 
consisting of just two projectors, 
is sufficient to optimally estimate $p$ and has been performed for several noise degrees.
For each level of noise, we estimated the channel parameter $p_{exp}$ as
$p_{exp}=\frac{\mathcal N_{ss}}{\mathcal N_{ss}+C_{int}}$
where $C_{int}$ are the coincidences between the two outputs of the BS
in interference condition and $\mathcal N_{ss}$ is the number of events in which
the two photons are detected on the same BS output side. $\mathcal N_{ss}$ was estimated
by knowing the amount of coincidences out of interference. 
The typical peak interference measured for the state $\ket{\psi^-}$ as a  function of the 
path delay $\Delta x$ is shown in Fig.\ref{piccobuca} of the Supplemental Material.
In Fig.\ref{probability}a) we report the experimental values $p_{exp}$ corresponding to
the different values of $T$.
In the corresponding inset we show the $p_{exp}$ errors evaluated by propagating the
$C_{int}$ and $\mathcal N_{ss}$ Poissonian errors.
They are in good agreement with the expected theoretical behavior.


{\it Ancillary assisted quantum process tomography.- }
The experimental results, just discussed for the optimal estimation
of the depolarizing channel, have been compared with the probability values of $p$
which can be obtained by exploiting the Ancillary assisted quantum process tomography (AAQPT)
\cite{DL,Alteprl03,mohs08pra,karjosab04}.
The action of a generic channel operating on a single qubit can be written as
$\mathcal E[\rho]=\sum^{3}_{i,j=0}\chi_{ij}\sigma_i\rho\sigma_j$,
where the matrix $\chi_{ij}$ characterizes completely the process.

AAQPT is based on the following procedure: i) prepare a two-qubit maximally entangled state and reconstruct it
by Quantum state tomography (QST) \cite {jam01pra};
ii) send one of the two entangled qubits through the channel $\mathcal E$; iii) reconstruct the output
two-qubit state by QST and obtain, in this way, the matrix $\chi_{ij}$ from the two-qubit output density matrix.
For a DC,
the matrix $\chi_{ij}$ is expressed as \cite{nielsen}
\begin{equation}
{\mathcal \chi^{Theo}_p}=
\begin{pmatrix}
(1-p) & 0  &  0  &  0\\

0  &\frac{p}{3}  &  0  &  0\\

0  &  0  &  \frac{p}{3}  &  0\\

0  &  0  &  0  &  \frac{p}{3}

\end{pmatrix}
\end{equation}


We implemented the AAQPT algorithm by injecting the state $\ket{\psi^-}$ into the DC and we
reconstructed by QST the density matrices of the input and output states for several noise degrees [see Fig.\ref{setup}a)].
We obtained the experimental matrix $\chi_{exp}$ for different values of $T$ and,
for each value of $T$, we found the parameter $p$ maximizing the fidelity between
the experimental $\chi_{exp}$ and the theoretical $\chi^{Theo}_p$ process matrices.
The experimental results are shown in Fig.\ref{probability}b). Even in this case the theoretical behavior is fully satisfied.
However, comparing these results with those obtained by the optimal
protocol, we observe that the latter leads to the same results, but with a much lower number of measurements.
In fact, in this case, only the two projections $M'$ are needed while, 
to implement the AAQPT algorithm, 16 measurements are necessary.
Moreover, by adopting our experimental setup we were able to demonstrate that the value
of $p$ and the DC action do not depend on the input state.
In fact the AAQPT was realized with all the four Bell states entering the DC, obtaining the same results of those
shown in Fig. \ref{probability}b).

It is worth noting that, even if the AAQPT gives a more complete information on the process
compared to the implemented optimal protocol, the latter
allows us to achieve a more accurate value of $p$.
The inset in Fig.\ref{probability}(a) shows that, for the optimal protocol, the measured standard deviation reaches the lower bound
given by $\sqrt{p(1-p)/N}$
[i.e. the square root of \eq{covmat} for the one-dimensional case divided by $N$], where
$N$ is the dimension of the sample used to evaluate $p$, thus demonstrating experimentally the 
attainability of the Cram\'er-Rao bound.
We show in the inset in Fig.\ref{probability}(b) the standard deviations, well above the optimal bound, obtained with the AAQPT (see the Supplemental Material 
for details about the numerical estimation).
The lower optimal bound represented by the black curve is below the experimental data, 
demonstrating that AAQPT is far away from the optimal 
estimation protocol presented in this work.

%

{\it Conclusion.- }
An optimal protocol allowing the most efficient
estimation of a noisy Pauli channel has been
experimentally implemented in this work.
The action of the noisy channel was introduced on one qubit of a maximally
entangled pair in a controlled way.
The efficiency of this method
has been compared to the one achieved by quantum process tomography,
demonstrating that the optimal protocol allows us to achieve the theoretical
lower bound for the errors and to perform the estimate of the noisy channel
with a lower number of measurements. This method can be profitably applied
when some knowledge on the noise process is available and
can be successfully implemented in quantum-enhanced technologies involving
the management of decoherence.

This work was supported by EU-Projects CHISTERA-QUASAR and CORNER, 
by the CNR project ''Hilbert``, and by the FARI project 2010 of Sapienza {Universit\`a} di Roma. GV was partially supported
by the Strategic-Research-Project QUINTET of the Department of
Information Engineering, University of Padova and the
Strategic-Research-Project QUANTUMFUTURE of the
University of Padova. SP was supported by CNPq, FAPEMIG, INCT-quantum information and Italian-Brazilian Contract CNR-CNPq (Quantum information
in a high dimension Hilbert space).

\section{Supplementary Information}
Ancillary Assisted Quantum Process Tomography (AAQPT) standard deviations, shown in the inset in Fig.3b) of the main text, have been calculated by 
a simulation realized by MATHEMATICA 5.0.
It was based on the following procedure:
\begin{itemize}
 \item a poissonian uncertainty was associated to the coincidence counts 
 \item hundred different matrices of the process were simulated 
 \item the fidelity between these matrices and the theoretical ones, reported in the main text, was calculated
 \item the routine NMAXIMIZE allowed to find numerically the value of $p$ maximizing the Fidelity for each simulated process
 \item this sample, composed of hundred values of $p$, was used to calculate both average and standard deviation.  
\end{itemize}

The experimental values of the inset in Fig.3b) correspond to 
a sample of $\sim$1600 coincidences per second.
It can be evaluated that a number of $150000$ coincidences per second, 
which is nearly two orders of magnitude larger, is needed to approach the standard 
deviation value obtained by the optimal protocol (i.e. the inset in Fig.3a) of the main text).


\newpage

\begin{figure}
\includegraphics[angle=270,width=8cm]{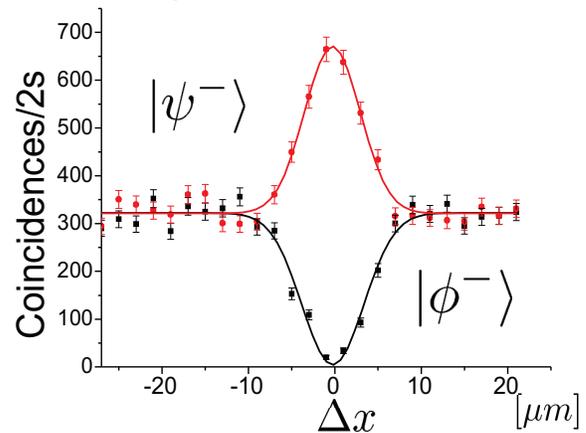}
\caption{Dip (peak) of the measured coincidence counts as a function of the optical path delay for a state $\ket{\phi^{-}}$($\ket{\psi^{-}}$) 
entering the BS in absence of noise.}
\label{piccobuca}
\end{figure}

\end{document}